\documentstyle[pre,aps]{revtex}

\input{epsf}

\def\beq{\begin{equation}}
\def\eeq{\end{equation}}

\begin{document}

\title      { Parameter estimation in nonlinear stochastic differential
                    equations
                }
\author {       J. Timmer\\
                Fakult\"at f\"ur Physik,
                Hermann -- Herder -- Str. 3,
                79104 Freiburg, Germany \\
                and \\
                Freiburger Zentrum f\"ur Datenanalyse und Modellbildung,
                Eckerstr. 1,
                79104 Freiburg,
	        Germany \\
		e-mail: jeti@fdm.uni-freiburg.de\\
		Phone: +49/761/203-5829 \\
		FAX: +49/761/203-5967\\
                         }
\date {\today }
\maketitle
\begin {abstract}
We discuss the problem of parameter estimation in nonlinear
stochastic differential equations based on sampled time series.
A central message from the theory 
of integrating stochastic differential equations is that there exists 
in general two time scales, i.e.~that of integrating these equations and that
of sampling. We argue that therefore maximum likelihood
estimation is computational extremely expensive. We discuss
the relation between maximum likelihood and quasi maximum 
likelihood estimation. In a simulation study,
we compare the quasi maximum likelihood method with an approach for 
parameter estimation in nonlinear stochastic differential
equations that disregards the existence of the two time scales.

\end {abstract}



%\clearpage

\twocolumn

\section {Introduction}

The analysis of complex systems by nonlinear 
deterministic differential equations has attracted much 
attention in recent years. Given a parameterized
differential equation, best-fit parameters can be obtained by
a least-squares minimization, see 
e.g.~\cite{bock83,gouesbet91,baake92a,baake92b,timmer98d}.

Often, however, the dynamics does not follow a strict deterministic law. 
In dissipative systems, due to fluctuation-dissipation theorems, 
thermal noise might have to be taken into account \cite{keizer93}.
Furthermore, in open nonequilibrium systems like in biology 
and economy, the dynamics is often also exposed to highdimensional, 
effectively random, influences, see \cite{timmer98d} for an example. 
This calls for a description by nonlinear stochastic differential 
equations (SDE), respectively their corresponding Fokker-Planck equations
\cite{gardiner97,vankampen85,honerkamp_buch}.
While parameter estimation in linear stochastic and nonlinear deterministic
differential equations even for data covered by additive 
observational noise is well known, see \cite{timmer98d} for a review, 
the estimation of parameters in nonlinear SDEs is still under development.
In this paper we discuss three approaches for parameter estimation
in nonlinear SDEs. The results 
carry over to modeling by Fokker-Planck equations.

The organization of this paper is as follows: In the next section, we 
briefly summarize the theory of integrating SDEs and exemplify its
practice for the  van der Pol oscillator undergoing stochastic forcing. 
In Section \ref{estimation} we discuss three approaches for parameter 
estimation in nonlinear stochastic differential equations.
In Section \ref{simulations} we compare two of these approaches
in a simulation study using the stochastic van der Pol oscillator.

\section{Integrating nonlinear stochastic differential equations} 
			\label{integration}

\subsection {Theory}

A stochastic differential equation (SDE) with parameter vector 
$ \vec{\theta} $ is given by:
\beq \label{sde_general}
     \dot{\vec{x}} = \vec{a}(\vec{\theta},\vec{x}) + 
		\underline{b}(\vec{\theta},\vec{x}) 
				\, \vec{\epsilon} \quad ,
\eeq
where $ \vec{\epsilon} $ denotes uncorrelated Gaussian noise
or, in mathematical terms, the increment of Brownian motion.
This general form includes additive and multiplicative 
dynamical noise. $\vec{a}(\vec{\theta},\vec{x})$ is usually denoted
as (deterministic) drift term, $\underline{b}(\vec{\theta},\vec{x})$ 
is called the diffusion term.

The integration of a SDE is not straightforward.
This is due to the mathematical problems of evaluating integrals
which involves the dynamical noise $\vec{\epsilon}$, see \cite{kloeden91} 
for a brief introduction and \cite{kloeden92} for a detailed discussion.
Applying the same ideas underlying higher order integration schemes
for deterministic differential equation like Runge-Kutta \cite{recipes}
to SDEs leads to hardly treatable stochastic integrals given in
\cite{honerkamp_buch}. Thus, only low order integration schemes
can be used. The lowest order so-called Euler-scheme for 
Eq.~(\ref{sde_general}) is given by :
\beq \label{dicr_sde}
\vec{x}(t+\delta t) = \vec{x}(t) + \delta t\, \vec{a}(\vec{\theta},\vec{x}(t)) +
     \sqrt{\delta t}\, \underline{b}(\vec{\theta},\vec{x}(t)) \vec{\epsilon}(t) 
			\quad ,
\eeq
which is of order $1/2$ for multiplicative noise and of order $1$ for
additive noise. The characteristics of integration 
schemes for SDEs is the $ \sqrt{\delta t} $ term which results from 
the integration rules for white noise \cite{kloeden92}.

For the task of parameter estimation, we assume that the system under 
observation can be adequately
described by Eq.~(\ref{sde_general}) with unknown parameter vector
$ \vec{\theta} $. The process is observed at discrete time points
given by the sampling interval $\Delta t$.
For nonlinear deterministic differential equations it is usually
possible to chose identical time steps for the integration and the sampling.
The necessity of using a low order scheme for SDEs means that the 
integration step size $\delta t $ is usually much smaller than the sampling 
interval $\Delta t$ by which the time series is recorded. 
Thus, while Eq.~(\ref{dicr_sde}) can be written as
\beq \label{eq_h}
  \vec{x}(t+\delta t) = \vec{h}(\vec{x}(t)) + \nu(t) \quad ,
\eeq
with $ \vec{h}(\vec{x}(t)) $ a function that can be related back to the 
parameter vector $ \vec{\theta}$ of 
$ \vec{a}(\vec{\theta},\vec{x}(t)) $ and $\nu(t)$ uncorrelated Gaussian noise,
on the time scale $ \Delta t $ of sampling the relationships are more 
intricate. While it is still possible to formulate the dependence
between present and future states as
\beq \label{eq_g}
  \vec{x}(t+\Delta t) = \vec{g}(\vec{x}(t)) + \eta(t) \quad ,
\eeq
the parameter vector $\vec{\theta}$ of $ \vec{a}(\vec{\theta},\vec{x}(t)) $
can, in general, not be inferred from $\vec{g}(\vec{x}(t))$.
Furthermore, $\eta(t)$, while still uncorrelated
with zero mean, does in general not represent Gaussian noise.
In other words, in nonlinear SDEs the relation between the mean 
of the conditional density $p(\vec{x}(t+\Delta t) | \vec{x}(t))$ 
and the drift-term $ \vec{a}(\vec{\theta},\vec{x}(t)) $ is
not explicitly known (The analogous problem is given for modeling 
such systems by their corresponding Fokker-Planck equation).
Furthermore, the conditional density $p(\vec{x}(t+\Delta t) | \vec{x}(t))$ 
is, in general, not Gaussian although $\vec{\epsilon}(t)$ in 
Eqs.~(\ref{sde_general},\ref{dicr_sde}) is Gaussian. For these
reasons parameter estimation in SDEs %iscussed in Section \ref{estimation} 
is a non-trivial task.

\subsection{An example} \label{example}

To exemplify the practical issues of integrating SDEs, we choose 
 the van der Pol oscillator \cite{vanderpol22}:
\beq \label{vdp}
\ddot{x} = \mu\,(1-x^2)\, \dot{x} - x,  \qquad \mu > 0.
\eeq

This system exhibits a limit cycle due to the amplitude dependent change 
of the sign of the damping term. We expose it to a random force of
unit variance, leading to 
\begin{eqnarray}
\dot{x}_1 & = & x_2  \label{sys1} \\
\dot{x}_2 & = & \mu\,(1-x_1^2)\, x_2 - x_1 + \epsilon  \label{sys2}\quad ,
\end{eqnarray}
where $ x_1$ denotes the location and  $ x_2 $ the velocity.
The Euler integration scheme for Eqs.~(\ref{sys1},\ref{sys2}) is
given by
\begin{eqnarray}
x_1(t+\delta t) & = & x_1(t) + \delta t\, x_2(t) \label{discr1}\\
x_2(t+\delta t) & = & x_2(t) + \delta t\, (\mu\,(1-x_1^2(t))\, x_2(t) - x_1(t)) 
	+ \sqrt{\delta t}\, \epsilon(t) \quad . \label{discr2} 
\end{eqnarray}

In the following we chose $\mu=3$.
To obtain a sampled time series of the system, one has to decide
on $\delta t$ and $\Delta t$. $\Delta t$ should be chosen so
that the process is reasonably sampled.
Characteristics of the stochastic van der Pol oscillator
and related perturbed limit cycles have been investigated
in \cite{kurrer91,leung95}. These authors have shown that the mean
period of this system is slightly smaller than the period
of the corresponding deterministic system which is approximately 9 s.
By choosing $\Delta t=0.5$, we obtain approximately 18 data
points per mean period. 

The choice of the integration step size  
$\delta t$ is more difficult. For deterministic systems
adaptive algorithms are well established that guarantee 
an upper bound for the deviation from the true trajectory \cite{recipes}.
No corresponding straightforward procedure is available for SDEs.
For SDEs the characteristic quantities are the conditional densities 
$p(\vec{x}(t+\Delta t) | \vec{x}(t))$. To obtain a sampled
trajectory that can be regarded as a realization of the
SDE, $\delta t$ has to be chosen such small that the 
conditional densities become independent of  $\delta t$
for smaller values. Figs.~\ref{cond_dens1} and \ref{cond_dens2}
shows this procedure for the first component of
$\vec{x}(t+\Delta t)$ of the stochastic van der Pol oscillator 
denoted by $x_1(t+\Delta t)$ for the two state vectors 
$\vec{x}(t)=(-0.0935,-4.284) $ and 
$\vec{x}(t)=(1.021,-0.9375) $.  
%
% \marginpar {Please locate Fig.~\ref{cond_dens1} and Fig.~\ref{cond_dens2}
% around here}
%
The conditional densities 
$p(x_1(t+\Delta t) | \vec{x}_i(t))$ were estimated
by triangular kernel estimators \cite{silverman} 
based on 5000 integrations of the SDE. The quality of these
estimated densities with respect to bias and variance depends
on the width of the kernel. By visual inspection, the width of the kernel
was chosen
equal to 0.02 in  Fig.~\ref{cond_dens1} and equal to 0.1
in Fig.~\ref{cond_dens2}.
In Fig.~\ref{cond_dens1}, the estimated conditional density 
changes drastically between $\delta t=0.1 s$ and $\delta t=0.01 s$.
It becomes
independent from  $\delta t$ for $\delta t=0.001s $. 
In Fig.~\ref{cond_dens2}, this already happens for $\delta t=0.01 s$. 
The reason for the different behavior is that the time evolution
in the first case experiences more of the nonlinearity
of the system. By investigating numerous analogous simulations,
we regarded  $\delta t=0.001 s$ as an appropriate integration
time step. The two state space vectors $\vec{x}(t)$
 used for the above simulation were
realizations of the trajectory obtained with $\delta t=0.001 s$.
Thus, the procedure for determining $\delta t$ is an interactive one that
has to be selfconsistent.

Fig.~\ref{data} shows a realization of the stochastic van der Pol
oscillator with $\mu=3$, $\delta t= 0.001$ and  $\Delta t= 0.5$.
%
%\marginpar {Please locate Fig.~\ref{data} around here }
%
The two data points marked by arrows at time 28s, respectively
28.5s were used for the estimation of the conditional 
densities shown in Figs.~\ref{cond_dens1} and \ref{cond_dens2}.

%Fig.~\ref{data_phase} displays the data in the phase space.

\subsection {A consequence}

The above considerations also have consequences for 
mathematical models used in simulation studies and 
proposed for analyzing time series
in the frame of nonlinear stochastic systems by difference
equation of the type :
\beq
x(t+1) = f (\vec{p}, x(t), x(t-1), \ldots, x(t-m)) + \epsilon(t) \quad .
\eeq
Here the sampling interval is set to unity, $ \epsilon(t) $ denotes 
uncorrelated Gaussian noise, $\vec{p}$ the parameters and $m$ the 
order of the model, 
see e.g.~\cite {tong96}. For these models the conditional distribution
of $x(t+1)$ given the history is Gaussian.
Due to the Gaussianity, least squares optimization leads
to a maximum likelihood estimation of the parameters which is 
asymptotically unbiased and efficient \cite{lehman83,azzalini96}.
Figs.~\ref{cond_dens1} and \ref{cond_dens2} show that for difference
equations that are thought to be discretized versions of SDEs,
the dynamical noise should be non-Gaussian, see the skewed distribution
in Fig.~\ref{cond_dens2}, and state dependent
heteroscadistic. It might even be multimodal.

\section{Parameter estimation} \label {estimation}

In this section we discuss three methods to estimate parameters
in SDEs. Due to its superior statistical properties, the most
desirable method would be a maximum likelihood estimation
\cite{lehman83,azzalini96}.
We argue that this approach is not feasible.
Then we discuss recently suggested quasi-maximum likelihood
approaches \cite{bibby95}. 
The third approach applies the integration scheme, 
Eq.~(\ref{dicr_sde}), for the whole sampling interval by using 
$\Delta t = \delta t$. With respect to the identification of
the two time scales this last approach is similar to a
procedure to estimate parameters in SDEs 
proposed in \cite{borland92a,borland92b,siegert98}.
In the simulation studies presented
in these publications the time series were sampled
at the time step of integration.

To simplify the exposition, we consider a scalar dynamics and
a single parameter $\theta$ in the following.

\subsection{Maximum likelihood estimation}

Denoting the stationary distribution of the SDE for a given parameter
$\theta$ by $ \pi(x|\theta) $,
the likelihood for a sampled time series of length $N$ reads :
\beq
{\cal{L}}(x(t_1), x(t_2), \ldots, x(t_N);\theta)= \pi(x(t_1)|\theta)
\prod_{i=1}^{N-1} p(x(t_{i+1}) | x(t_i),\theta) \quad .
\eeq

Maximizing ${\cal{L}}(.;\theta)$ leads to an estimator $ \hat{\theta} $ that 
is asymptotically unbiased and has least conservative confidence regions.
Note that a biased estimator can lead to erroneous interpretations
of results, while
suboptimal confidence regions 'only' lowers the power to reject
simpler models in favor of more complex ones, but does not lead 
to statistically false positive results.

Usually, the logarithm of the likelihood is considered and the
term $ \pi(x(t_1)|\theta) $ whose influence vanishes asymptotically
is neglected:
\beq \label{loglik}
L(x(t_2), x(t_3), \ldots, x(t_N)|\theta)= 
\sum_{i=1}^{N-1} \ln \, p(x(t_{i+1}) | x(t_i),\theta) \quad .
\eeq
The maximum likelihood estimator $ \hat{\theta}_{mle} $ is defined by:
\beq \label {mle}
 \frac{d}{d \, \theta } L(x(t_2), x(t_3), \ldots, x(t_N);\theta)= 0 \quad .
\eeq

To obtain $ \hat{\theta}_{mle} $ an iterative optimization
strategy has to be applied. Starting from an initial guess
for the parameter, the conditional densities 
$  p(x(t_{i+1}) | x(t_i),\theta) $ in Eq.~(\ref{loglik})
have to be estimated and evaluated. This can be achieved by
solving the corresponding Fokker-Planck equation or by
applying the integration scheme, Eq.~(\ref{dicr_sde}),
with the actual guess of the parameters several times on the
intervals $[t_{i}, t_{i+1}]$ starting from the observed $ x(t_{i}) $
to estimate the conditional density, e.g.~by kernel estimation
\cite{silverman}. Then, the log-likelihood can be calculated.
Based on this procedure, the parameter is changed until
the log-likelihood is extremal by applying some optimization
algorithm \cite{recipes}. 
The performance of this procedure depends heavily on the
quality of the density estimation. Thus, for the procedure based
on integrating the SDE thousands
of trajectories in each interval  $[t_{i_1}, t_{i}]$
have to be realized. Note that the densities shown in 
Figs.~\ref{cond_dens1} and \ref{cond_dens2} that were based on
5000 realizations are not smooth enough to
enable a numerically stable estimation of the parameters.
 Furthermore, this procedure involves the choice
of a parameter determining the bandwidth of the density
estimator. This parameter has to be chosen data driven and
state dependent, e.g.~by a computer intensive cross-validation
procedure \cite{silverman,gasser79,craven79}. Therefore, the desirable 
method of maximum likelihood is hardly applicable.

\subsection{Quasi maximum likelihood estimation} \label {qmle}

Bibby and Sorensen \cite{bibby95} suggested
to use a quasi maximum likelihood estimator instead
of the infeasible maximum likelihood estimator, Eq.~(\ref{mle}).
The key idea of this procedure is that Eq.~(\ref{mle}) can 
formally be read as a search for a root :
\beq \label{est_eq}
  G(\theta) = 0 \quad ,
\eeq
defining an estimator $ \hat{\theta} $. By virtue of choosing
$G(\theta)$, the resulting computationally feasible 
estimator can be forced to be
unbiased by paying the price that the resulting confidence
region are no longer optimal. Experience shows that
the loss in optimality is often rather small \cite{heyde97}.
Eq.~(\ref{est_eq}) is called estimation equation, and the
resulting estimator is called quasi maximum likelihood estimator.

Bibby and Sorensen \cite{bibby95} have shown that in the case
of SDEs one possible choice for $G(\theta)$ is
given by:
\beq \label{est_eq_sde1}
  G(\theta) = \sum_{i=1}^{N-1} \,g_{i}(.)\, 
	(x(t_{i+1}) - E(x(t_{i+1})|x(t_{i}),\theta)) \quad ,
\eeq

where $g_i(.)$ is a function that is derived from the SDE and
$E(x(t_{i+1})|x(t_{i}), \theta)$ denotes the expected value of $ x(t_{i+1}) $
conditioned on the observed $ x(t_{i}) $ for a given $ \theta $.
Different possible procedures for deriving $g_i(.)$ given the SDE
are discussed in  \cite{bibby95}.

In opposite to the conditional density necessary for the maximum 
likelihood case, the conditional expectation in Eq.~(\ref{est_eq_sde1}) 
can the estimated
reliably with a comparable small numbers of integrations
in the intervals $[t_{i}, t_{i+1}]$ starting from the observed $ x(t_{i}) $.

\subsection{ ''$\Delta t = \delta t$'' approach} \label{delta_approach}

In the third approach we use the discretization scheme, 
Eq.~(\ref{dicr_sde}), on the interval of the whole sampling time step $\Delta t$
by setting $\Delta t = \delta t$ in Eq.~(\ref{dicr_sde}). This yields
an estimate  $\widehat{x}((t_{i+1})|x(t_{i}),\theta)$.
The parameters are adjusted  until the mean square error
\beq
     \sum_{i=1}^{N-1} (x(t_{i+1}) - \widehat{x}((t_{i+1})|x(t_{i}),\theta))^2
\eeq
 is minimized. 
This method neglects the existence of the two time scales for integrating and 
sampling SDEs.  Furthermore, it implicitly assumes that the variance of the
conditional density is state space independent and
Gaussian.

\section {Simulation study} \label{simulations}

We investigate the behavior of the quasi maximum likelihood and
the ''$\Delta t = \delta t$'' approach in a simulation study using 
the stochastic van der Pol oscillator introduced in 
Section \ref{example}. We integrated the process with $\delta t=0.001s$.
The smaller the sampling time, the smaller the differences between 
$\vec{h}(.)$ in Eq.~(\ref{eq_h}) and  $\vec{g}(.)$ in Eq.~(\ref{eq_g}) 
will be. Thus, the less severe the effects of ignoring the two time 
scales by the ''$\Delta t = \delta t$'' approach will be 
\cite{dacunha86,florens_zmirou89,shoji97}.
Therefore, we investigate the behavior of the two approaches
described above for different sampling times between $\Delta t=0.005s$ 
and $\Delta t=0.5s$.
As in the simulation study presented in \cite{siegert98}, we 
assume that the whole state space vector is observed.

For the quasi maximum likelihood approach, Section \ref{qmle},
we chose the term  $g_i(.)$ in Eq.~(\ref{est_eq_sde1}) to be:
\beq
   g_i(\vec{x}(t_{i})) = (1-x_1^2)\,x_2 \quad, 
\eeq
generalizing the line of argument of \cite{bibby95}. 
The term ''$(x(t_{i+1}) - E(x(t_{i+1})|x(t_i), \theta)$'' 
in Eq.~(\ref{est_eq_sde1}) is read as 
$(x_1(t_{i+1}) - E(x_1(t_{i+1})|\vec{x}(t_{i}), \theta)$. Thus, we
predict the observed location $x_1(t_{i+1})$ based on the whole state 
space vector at the present time step $t_{i}$. The expected value 
of $x_1(t_{i+1})$ is estimated based on 50 integrations.
Finding the zero of Eq.~(\ref{est_eq_sde1}) is performed by routines
from \cite{recipes}.

For the ''$\Delta t = \delta t$'' approach, Section \ref{delta_approach},
we use the information from the whole state space from the history
and from the present state. The minimization is perform by routines 
from \cite{recipes}.

Fig.~\ref{fig_res_3} displays the results of the simulation study
for sampling time steps $\Delta t$ ranging from $0.005$ to $0.5$. 
%
%\marginpar {Please locate Fig.~\ref{fig_res_3} around here}
%
The 2$\sigma$ confidence
intervals were calculated from 50 repetitions. The length of the
time series for different sampling times were chosen such that 1000
 data points enters the estimation. 
The quasi maximum likelihood approach
yields unbiased results independent from the sampling interval.
For realistic sampling intervals the  ''$\Delta t = \delta t$'' approach
gives strongly biased results. Only for a sampling time of $\Delta t \le 0.025$
its estimate is consistent with the true parameter. This would
require to sample the process with approximately 360 points 
per period. Note that classical time series like the sunspot data
and the Canadian lynx data are sampled with 11, respectively 10 
points per period, see e.g.~\cite{brockwell87}.

Fig.~\ref{fig_res_5} shows the results analogous to Fig.~\ref{fig_res_3}
for nonlinearity parameter $\mu=5$. 
%
%\marginpar {Please locate Fig.~\ref{fig_res_5} around here}
%
The relative bias of 
the ''$\Delta t = \delta t$'' approach for
sampling intervals $\delta t= 0.05s, 0.1s $ and $ 0.5s$ is larger than 
for $\mu=3$ due to the higher degree of nonlinearity. 
Thus, in general, a sufficient sampling for the linearly approximating 
''$\Delta t = \delta t$'' approach to work depends on the degree
of nonlinearity.

\section {Discussion}

Modeling time series of open nonequilibrium systems by nonlinear 
stochastic differential equations (SDE) allows to take into account
the effects of the huge number of degrees of freedom that might be
active in such systems. 
A fundamental problem in estimating parameters in SDEs
is taught by the theory of integrating nonlinear stochastic 
differential equations. 
A central message from this theory
% of integrating nonlinear stochastic differential equations 
is that there are in general two time scales: 
that of integration and that of sampling. 
As a consequence, the mean and the distributional
characteristics of conditional densities on the time scale of
sampling can, in general, not be related back to the parameters of the
stochastic differential equation (SDE).

We discussed three approaches for parameter estimation
in SDEs based on sampled time series.
Unfortunately, the desirable maximum likelihood
approach is not feasible in the present context due to its
extreme computational burden to estimate the 
conditional distributions. It might become 
tractable with the advent of more powerful massive parallel computers.
In the presented simulation study, the quasi maximum likelihood approach
yielded unbiased estimates for the parameter.
Disregarding the existence of the two time scales of integration and 
of sampling in SDEs and using the discretization scheme of the SDE 
itself on the time scale of the sampling led to results that are
unbiased only for the case of heavy oversampling, but biased
for conventional sampling rates.

Thus, methods that require the sampling time interval to be an admissable 
integration time step, like suggested in \cite{borland92a,borland92b},
are applicable to measured time series only for sufficiently sampled data.
The same holds for the desirable approach to nonparametrically estimate the 
complete functional form of the deterministic drift term of
a stochastic differential equation based on the discretization 
of the corresponding Fokker-Planck equation proposed in \cite{siegert98}
where formally even the limit ''sampling interval $ \rightarrow 0$''
is required.

The approaches discussed in this paper require to observe the
complete state space vector which is usually not possible. Our future
work will concentrate on the generalization to the case
of parameter estimation in nonlinear stochastic differential equations 
based on scalar observations as it is already possible in linear
stochastic and nonlinear deterministic systems \cite{timmer98d}.

%\clearpage

%\bibliography{/u/honer/jeti/tex/paper/jetilit}
%\bibliographystyle{prsty}
%\bibliographystyle{plain}

%\clearpage

\section*{Figure captions}

\begin{figure}
\epsfxsize=8.4cm
\epsfbox{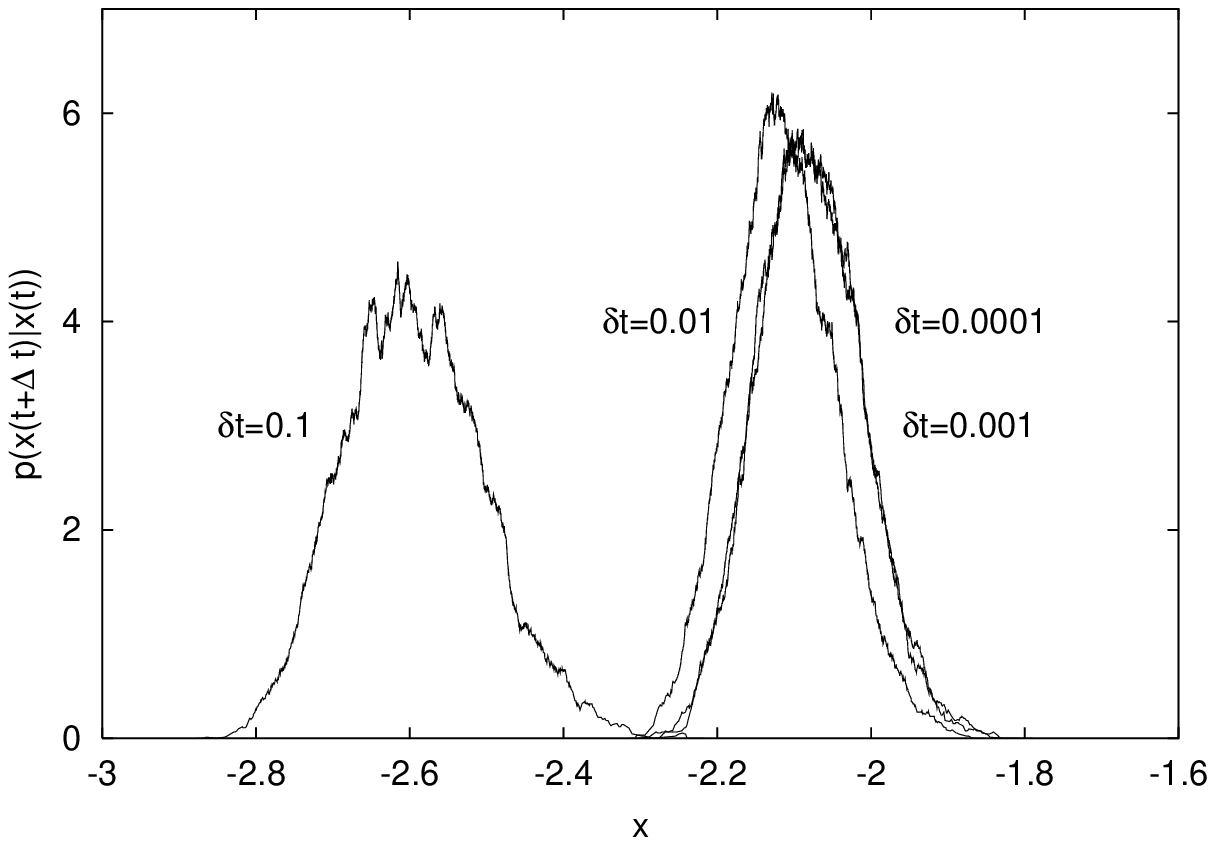}
\caption{}
\label{cond_dens1}
\end{figure}

\begin{figure}
\epsfxsize=8.4cm
\epsfbox{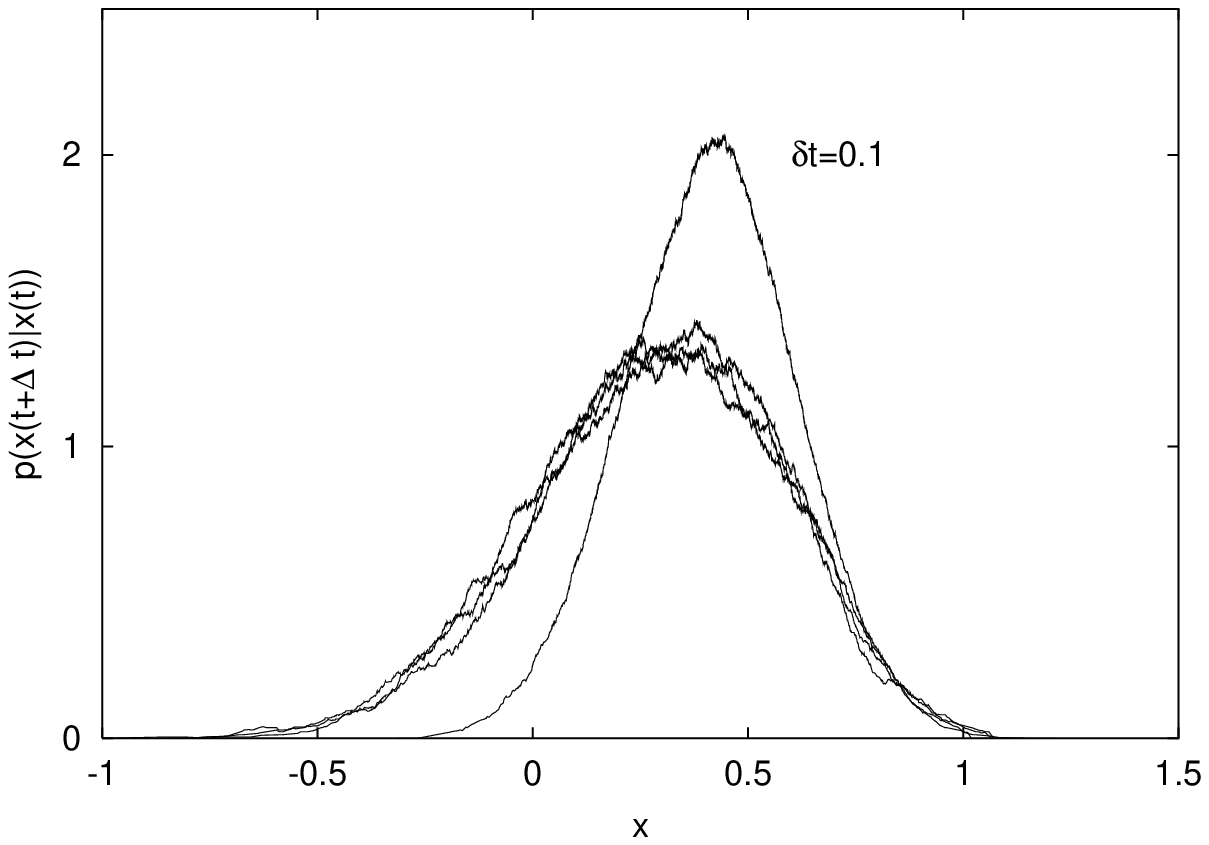}
\caption{}
\label{cond_dens2}
\end{figure}

\begin{figure}
\epsfxsize=8.4cm
\epsfbox{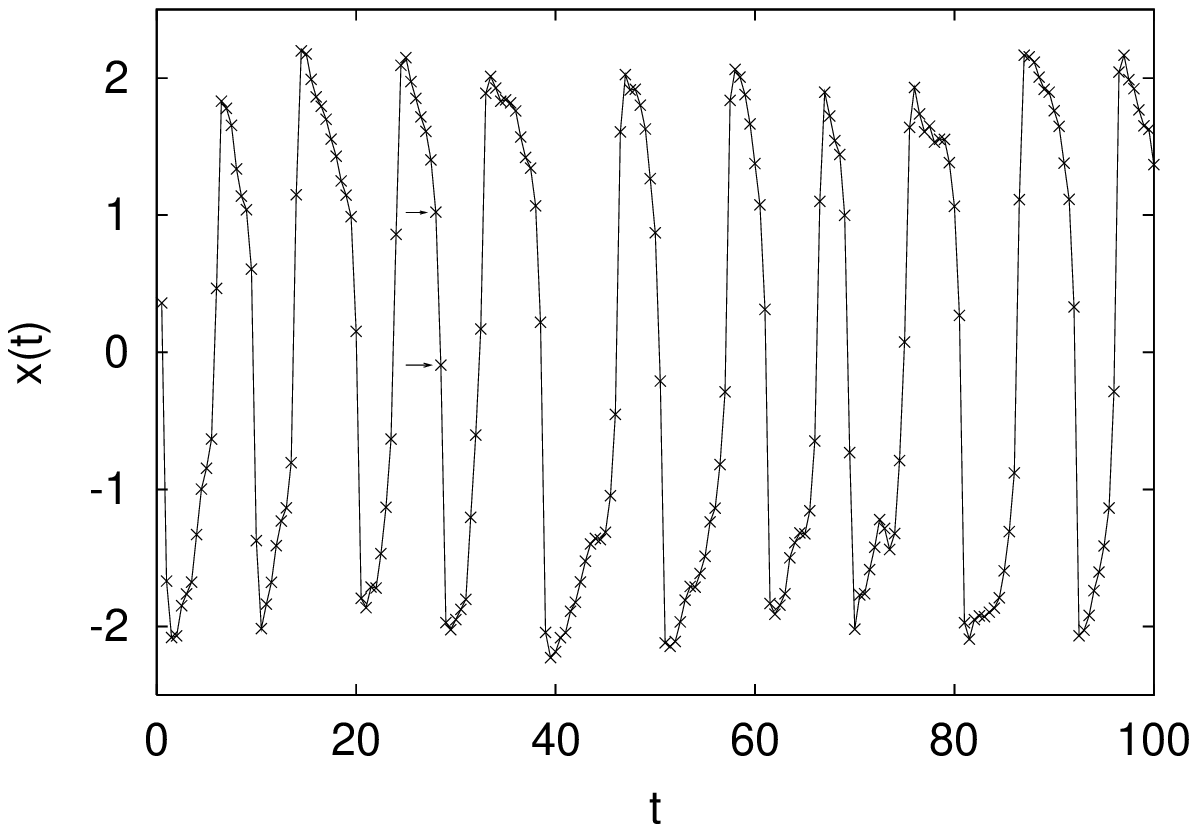}
\caption{}
\label{data}
\end{figure}

\begin{figure}
\epsfxsize=8.4cm
\epsfbox{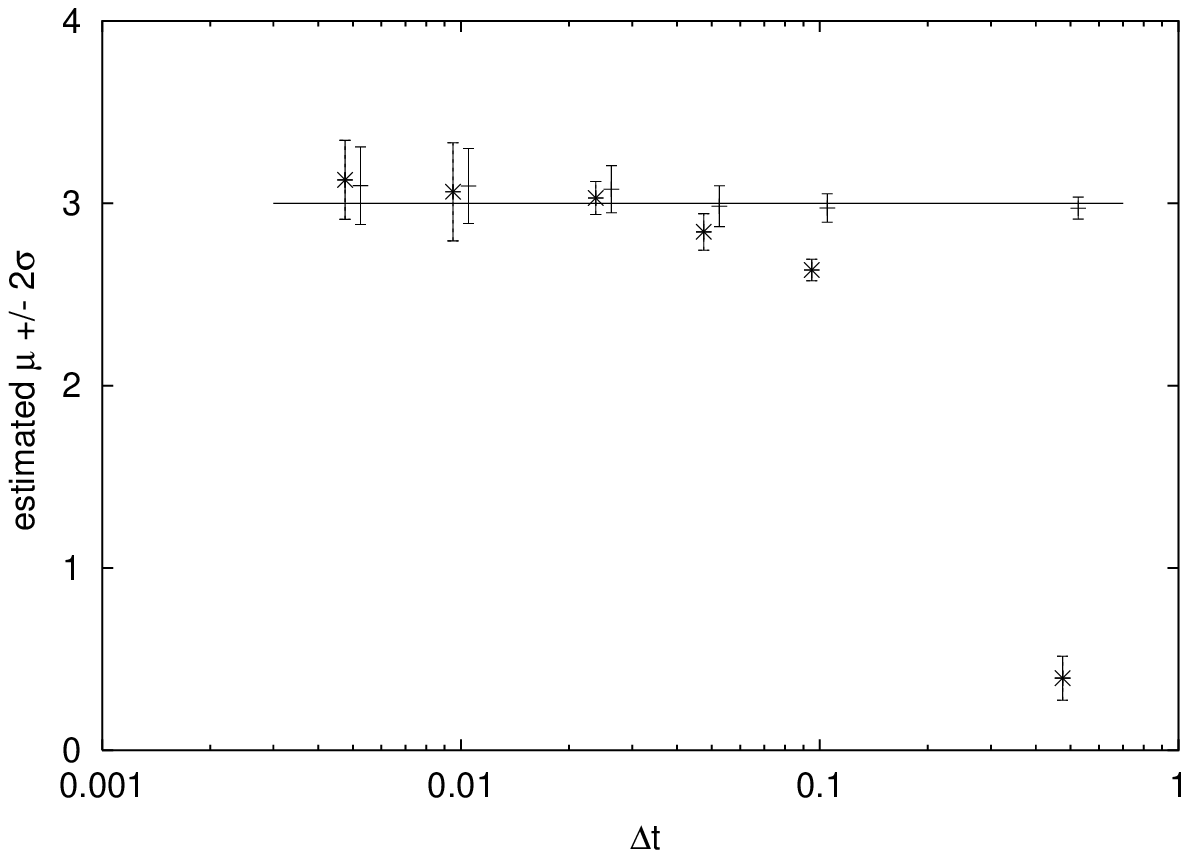}
\caption{}
\label{fig_res_3}
\end{figure}

\begin{figure}
\epsfxsize=8.4cm
\epsfbox{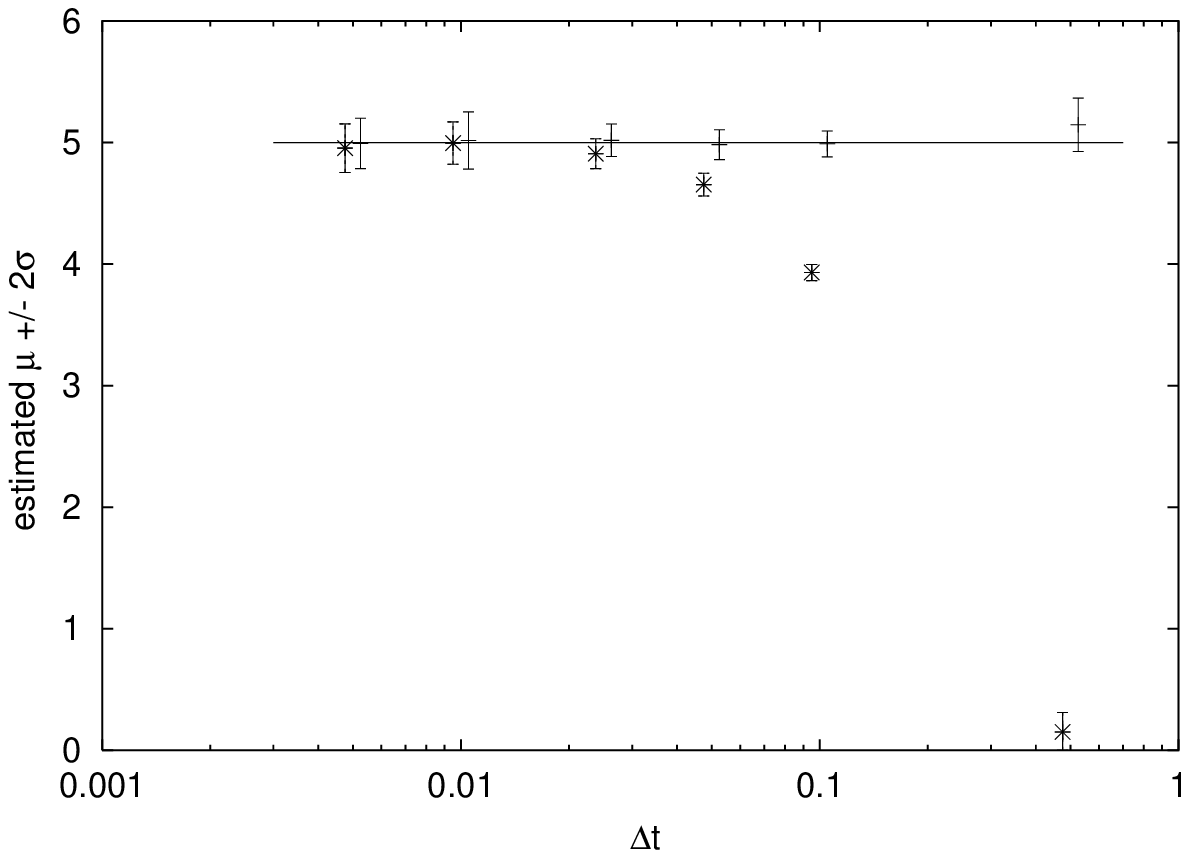}
\caption{}
\label{fig_res_5}
\end{figure}

\begin{itemize}
 
\item [Fig.~\ref{cond_dens1}]  Estimated conditional densities
 $p(x_1(t+\Delta t) | \vec{x}(t))$ 
for the stochastic van der Pol oscillator for $\mu=3$, 
$\vec{x}(t)= (-0.0935,-4.284)$,
integration time steps $\delta t=0.1s, 0.01s, 0.001s, 0.0001s$
and sampling interval $\Delta t=0.5s$. 

\item [Fig.~\ref{cond_dens2}] Estimated conditional densities
 $p(x_1(t+\Delta t) | \vec{x}(t))$ 
for the stochastic van der Pol oscillator for $\mu=3$, 
$\vec{x}(t)= (1.021,-0.9375)$,
integration time steps $\delta t=0.1s, 0.01s, 0.001s, 0.0001s$
and sampling interval $\Delta t=0.5s$. 
 
\item [Fig.~\ref{data}] Realization of the stochastic van der Pol 
oscillator for $\mu=3$, integration time step $\delta t=0.001s$
and sampling interval $\Delta t=0.5s$.
The two data points marked by arrows at time 28s, respectively
28.5s were used for the estimation of the conditional 
densities shown in Figs.~\ref{cond_dens1} and \ref{cond_dens2}.

\item [Fig.~\ref{fig_res_3}] Dependence of the estimated parameter 
$\hat{\mu}$ of the stochastic van der Pol oscillator on the sampling 
interval. The integration time step $\delta t $ was 
$0.001s$. The 95\% confidence intervals were calculated based on 50 
repetitions. +: quasi maximum likelihood approach, 
 $\ast$: ''$\Delta t = \delta t$'' approach.
For sake of clearity the results are slightly scattered around
the applied sampling time steps $\Delta t=0.005s, 0.01s, 0.025s, 0.05s,
0.1s $ and $0.5s$.
The true value of the parameter $\mu=3$ is marked by the solid line.  

\item [Fig.~\ref{fig_res_5}] Analogous to Fig.~\ref{fig_res_3} for $\mu=5$.

\end{itemize}

%\clearpage

%\clearpage

%\clearpage

%\clearpage

%\clearpage

\end {document}